\documentclass[reprint,
superscriptaddress,
amsmath,amssymb,
aps,
floatfix,
longbibliography]{revtex4-1}

\usepackage{amsmath}
\usepackage{graphicx}
\usepackage{multirow}
\usepackage{dcolumn}
\usepackage{bm}
\usepackage{xcolor}      

\begin{document}

\title{Polymer-free van der Waals assembly of 2D material heterostructures using muscovite crystals}

\author{Ian Babich}
\author{Timofey M. Savilov}
\author{Natalia A. Mamchik}
\affiliation{Institute for Functional Intelligent Materials, National University of Singapore,\\ 4 Science Drive 2, Singapore 117544, Singapore}
\affiliation{Department of Materials Science and Engineering,\\ National University of Singapore, Singapore 117575, Singapore}
\author{Kristina Vaklinova}
\affiliation{Institute for Functional Intelligent Materials, National University of Singapore,\\ 4 Science Drive 2, Singapore 117544, Singapore}
\author{Nansi Zhou}
\affiliation{Institute for Functional Intelligent Materials, National University of Singapore,\\ 4 Science Drive 2, Singapore 117544, Singapore}
\affiliation{Department of Materials Science and Engineering,\\ National University of Singapore, Singapore 117575, Singapore}
\author{Denis S. Baranov}
\affiliation{Institute for Functional Intelligent Materials, National University of Singapore,\\ 4 Science Drive 2, Singapore 117544, Singapore}
\author{Dmitrii A. Litvinov}
\affiliation{Institute for Functional Intelligent Materials, National University of Singapore,\\ 4 Science Drive 2, Singapore 117544, Singapore}
\affiliation{Department of Materials Science and Engineering,\\ National University of Singapore, Singapore 117575, Singapore}
\author{Virgil Gavriliuc}
\author{Yue Yuan}
\affiliation{Institute for Functional Intelligent Materials, National University of Singapore,\\ 4 Science Drive 2, Singapore 117544, Singapore}
\author{Amoz Chua}
\affiliation{Department of Materials Science and Engineering,\\ National University of Singapore, Singapore 117575, Singapore}
\author{Kenji Watanabe}
\author{Takashi Taniguchi}
\affiliation{Advanced Materials Laboratory, National Institute for Materials Science,\\ 1-1 Namiki, Tsukuba, 305-0044, Japan}
\author{Mario Lanza}
\author{Maciej Koperski}
\author{Kostya S. Novoselov}
\affiliation{Institute for Functional Intelligent Materials, National University of Singapore,\\ 4 Science Drive 2, Singapore 117544, Singapore}
\affiliation{Department of Materials Science and Engineering,\\ National University of Singapore, Singapore 117575, Singapore}
\author{Alexey I. Berdyugin}
\email{Corresponding authors: \\alexey@nus.edu.sg, m.siskins@soton.ac.uk}
\affiliation{Institute for Functional Intelligent Materials, National University of Singapore,\\ 4 Science Drive 2, Singapore 117544, Singapore}
\affiliation{Department of Materials Science and Engineering,\\ National University of Singapore, Singapore 117575, Singapore}
\affiliation{Department of Physics, Faculty of Science, National University of Singapore, Singapore 117551, Singapore}
\author{Makars \v{S}i\v{s}kins}
\email{Corresponding authors: \\alexey@nus.edu.sg, m.siskins@soton.ac.uk}
\affiliation{Institute for Functional Intelligent Materials, National University of Singapore,\\ 4 Science Drive 2, Singapore 117544, Singapore}
\affiliation{School of Physics and Astronomy, University of Southampton, Highfield, Southampton SO17 1BJ, United Kingdom}

\begin{abstract}
The advent of van der Waals (vdW) heterostructures has enabled formation of bespoke materials with atomic precision, where numerous quantum and topological phenomena have already been discovered. This atomic-layer tunability, however, comes at a cost: individual 2D layers must be picked up, moved, and placed in a deterministic manner while keeping their interfaces atomically clean. Recent advances in machine learning and robotics place even stronger emphasis on the deterministic aspect of vdW assembly. Current polymer-based transfer methods satisfy neither the determinism nor cleanliness requirements. To this end, solutions are needed where adhesion can be dynamically and deterministically controlled without leaving organic contamination. Here, we present a polymer-free transfer technique employing thin muscovite (mica) crystals. Temperature control over mica adhesion enables deterministic pick-up, stacking, and release of 2D materials, while their crystalline, inorganic nature ensures pristine interfaces and suppresses strain. Fully compatible with existing fabrication workflows, this approach enables the assembly of demanding vdW heterostructures, including those with exposed conductive layers, moir\'{e} superlattices and suspended membranes. Our method represents a promising strategy for vdW heterostructure fabrication toward its automatization.
\end{abstract}

\maketitle
\section*{Introduction}
Since the discovery of pristine two-dimensional (2D) materials, advances in nanofabrication have enabled the creation of precisely stacked assemblies of different atomically thin crystals, known as van der Waals (vdW) heterostructures \cite{HeteroGeim2013,HeteroNovoselov2016,HeteroLiang2019}. Over the past two decades, each advancement of fabrication procedures has not only improved the quality of existing heterostructures \cite{ReviewFrisenda2018,ReviewPham2024} but also has further expanded the design space for devices, allowing to observe previously inaccessible phenomena \cite{Martin2025}. These advancements have transformed condensed matter research by enabling heterostructures with tunable strong electronic correlations, a variety of symmetry-broken states at zero and high magnetic fields, and band reconstructions beyond what is achievable in pristine 2D materials \cite{Cao2018,Cao2018_2,Lu2019,Guo2025,Zondiner2020}. Yet, despite these achievements, there remains substantial room for further improvement in state-of-the-art assembly techniques \cite{ReviewPham2024,Martin2025}.

Early nanofabrication of 2D devices relied on liquid-assisted transfers \cite{WetReina2008,ReviewFrisenda2018, ReviewPham2024}, although not optimal for reliable assembly of mechanically exfoliated 2D materials into heterostructures \cite{ReviewFrisenda2018, Fan2020, ReviewPham2024}. The introduction of dry elastomer transfer with polymethyl-methacrylate (PMMA) \cite{PMMAReina2009,PMMADean2010} and  polydimethylsiloxane (PDMS) \cite{PDMSCastellanosGomez2014,PDMSYan2025} improved such fabrication significantly, increasing both yield and control of pick-and-place operations \cite{PMMADean2010,PDMSCastellanosGomez2014,PDMSYan2025}. However, PMMA and PDMS stamps, are viscoelastic: creep and relaxation during contact can induce uncontrolled, spatially varying strain, their surface can transfer residues, and adhesion is only coarsely tunable via contact time and temperature \cite{PDMSCastellanosGomez2014}. To gain more deterministic adhesion control, hot-polymer transfer methods using polycarbonate (PC) and polypropylene carbonate (PPC) became the workhorse for high-quality stacks \cite{PCZomer2014,PPCWang2013, PPCPizzocchero2016}. By tuning temperature through polymer glass transitions, one can pick up, rotate, and release a range of 2D materials with sub-micrometer precision \cite{PPCPizzocchero2016}. However, stacks are often left with persistent residues on the surface \cite{ReviewPham2024}, bubbles and blisters embedded into 2D material interfaces \cite{CleaningPurdie2018,Fan2020, BubbleIwasaki2020}, while the elevated temperatures used can cause layer slippage \cite{Annett2016} and alignment angle drift across large device areas \cite{Wang2016} with strain gradients that remain challenging to eliminate \cite{Fan2020,Tran2024,Martin2025}. Recently, state-of-the-art polymer-free strategies emerged, where freestanding 2D-material cantilevers \cite{CantileverJin2023,AdvMatLee2025} and flexible SiN$_x$/metal membranes coated with thin metals \cite{Manchester2023} are used as a stiffer transfer support, providing cleaner interfaces and surfaces for assembly of ultraclean and complex stacks \cite{Manchester2023}. These approaches mitigate residue but suffer from other implementation barriers: their performance depends on preparation of cantilevers (metal composition, deposition conditions, resulting surface roughness), which limits their applications, and the precise control and uniformity of twist angles over macroscopic areas relevant for opto-electronic studies have yet to be demonstrated \cite{Manchester2023, ReviewPham2024}.

\begin{figure*}
\begin{center}
	\includegraphics[width=\linewidth]{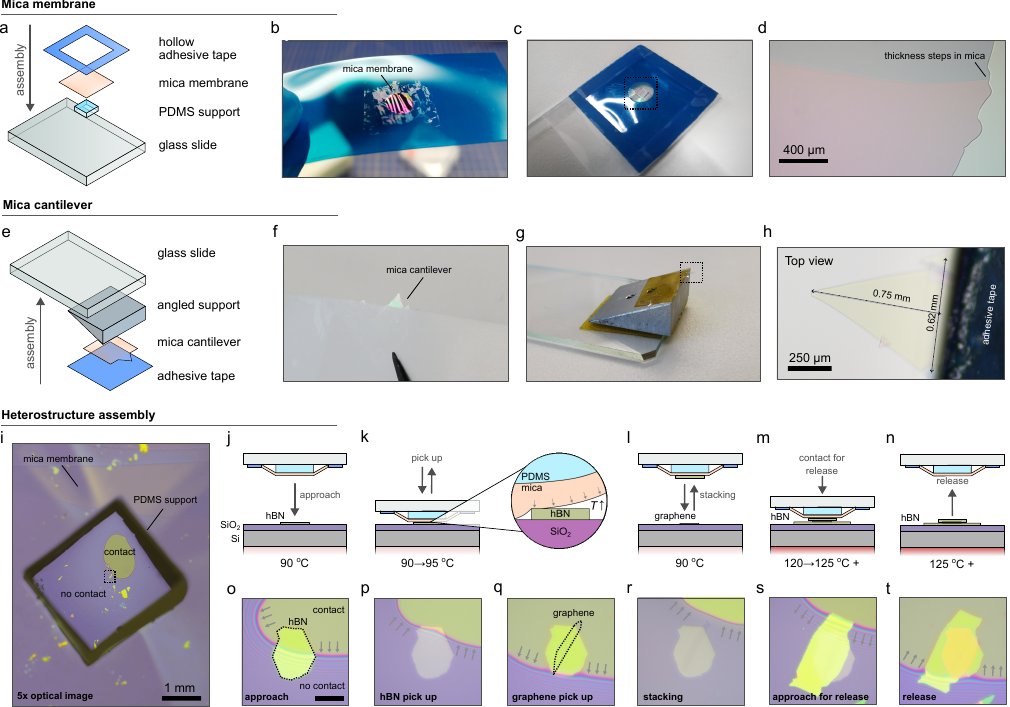}
	\caption{Standard working procedures for mica-assisted heterostructure assembly. \textbf{a-d} Preparation of the transfer stamp with polydimethylsiloxane (PDMS)-supported mica membrane. \textbf{a} Schematic diagram of the assembly process. \textbf{b} Photograph of the exfoliated freestanding mica membrane supported by the hollow adhesive tape. \textbf{c} The same tape transferred onto PDMS-on-a-glass slide support. \textbf{d} Optical image of the resulting mica membrane surface. \textbf{e-h} Preparation of mica cantilevers. \textbf{e} Schematic diagram of the assembly process. \textbf{f} Photograph of the exfoliated mica cantilever supported by the adhesive tape. \textbf{g} The same mica cantilever mounted on the angled support (35$^\circ$). \textbf{h} Optical image of the resulting mica cantilever surface. \textbf{i} Optical image of the typical heterostructure assembly process. \textbf{j-n} Step-by-step schematics of the heterostructure assembly with the mica membrane method, together with corresponding optical images shown in \textbf{o-t}. hBN - hexagonal boron nitride. Grey arrows indicate the direction of contact line movement. Black dashed lines indicate the outline of the relevant flake. Scale in \textbf{o-t} corresponds to scale bar in \textbf{o} of $40$ $\mu$m.}
	\label{Fig1}
\end{center}
\end{figure*}

Here, we introduce a deterministic, reliable, accessible and inexpensive all-dry polymer-free transfer methodology that utilizes muscovite (mica) crystals as stamps and cantilevers to stack van der Waals heterostructures. Mica combines three properties that directly target the long-standing limitations: first, this optically transparent and chemically inert van der Waals crystal cleaves with atomically flat and pristine surfaces on the millimeter scale; Second, its elastic stiffness and crystallinity suppress local deformations providing solid support and thereby minimizing mechanical stress for 2D material at any temperature during the transfer; Third, adhesion between mica and diverse 2D materials can be finely tuned using relatively small temperature change, enabling deterministic pick-up, precise angular placement, and reliable release. Moreover, we show that the compatibility of our approach with existing 2D material nanofabrication processes allows for seamless integration into current manufacturing protocols, facilitating the ease of prompt adoption of the method. As a result, we demonstrate complex heterostructure devices with perfect surfaces, interfaces and twist-angle integrity between 2D materials in a stack achieved with our method. We reach the limits in defect density and electronic device quality for magneto-transport, as well as demonstrate the high quality of large area moir\'{e} superlattices beyond its established geometries, including in its substrate-free suspended membranes.

\section*{Results}
\subsection*{Fabrication concept and protocol}
Standard 2D material transfer protocols typically rely on the viscoelastic properties of polymers, particularly close to their glass transition, their softness and ability to envelop the surface underneath and thus support 2D layers \cite{PCZomer2014,PPCPizzocchero2016,PDMSCastellanosGomez2014}. In contrast, we utilize the van der Waals nature and atomically flat surfaces of thin mica crystals \cite{dePoel2014} as a transfer medium. The key principle behind the mica-assisted assembly lies in the unique balance of van der Waals adhesion forces: the adhesion between mica and a given 2D material, e.g. graphene or hexagonal boron nitride (hBN), is typically stronger than that between the 2D material and the SiO$_2$ substrate, yet weaker than the adhesion to itself or between other 2D materials in a stack \cite{Sanchez2018}. This adhesion hierarchy allows using thin exfoliated mica crystals to deterministically pick up 2D flakes from SiO$_2$ substrates, stack and subsequently release them onto the surface of another flake.

For assembling vdW heterostructures, we use stamps similar to the conventional PDMS/PC design, but replace the PC layer with a large, thin mica sheet, as illustrated in Fig.~\ref{Fig1}. We use commercially available atomic force microscopy-grade muscovite crystals that have a low level of heavy ion doping and known stoichiometry, as confirmed using X-ray photoelectron spectroscopy (XPS, see Supplementary Note 1). High optical transparency in visible spectra and structural rigidness \cite{MicaCastellanosGomez2012} of mica allow us to exfoliate large and thin layers which are used to substitute PC and PPC membranes \cite{PPCPizzocchero2016, CleaningPurdie2018,BubbleIwasaki2020, PCZomer2014} (Fig.~\ref{Fig1}a-d), and ii) as well as free-standing cantilevers \cite{Manchester2023} (Fig.~\ref{Fig1}e-h) in conventional transfer methods. As a rule of thumb, we found the membrane thickness sufficient for successful transfer when thin film interference results in bright and distinctive colours ranging from red to purple, which correspond to approximately $150-650$ nm of mica (Fig.~\ref{Fig1}d and Supplementary Note 2). The resulting stamps and cantilevers are optically transparent and can be used for deterministic assembly of 2D material heterostructures under the optical microscope similar to conventional transfer techniques \cite{PDMSCastellanosGomez2014,PPCPizzocchero2016, PCZomer2014} (see Methods). This highlights the ease of integration of the mica-based method into existing 2D material manufacturing processes without any special adaptation of the transfer setups.

We empirically developed the standard recipe for hexagonal boron nitride (hBN) based heterostructure assembly, which relies on stronger adhesion of 2D materials to mica and each other than to a SiO$_2$ substrate \cite{Sanchez2018}. First, the Si/SiO$_2$ substrate with the hBN flake of interest is pre-heated to a range between $50-90$ $^{\circ}$C. The mica stamp, mounted on the transfer arm, is then manually brought in contact with a substrate by manipulating the $z$-axis (Fig.~\ref{Fig1}j). The contact area has a bright colour contrast to the rest of the substrate and is easily distinguishable under an optical microscope (Fig.~\ref{Fig1}i). Then in contact, the front line between the mica and the substrate can be controllably guided either by approach distance $z$ or by changing the substrate temperature in a range of $\pm10^{\circ}$ C (Fig.~\ref{Fig1}k and o). To successfully pick up the 2D material flake, mica should be in full contact with the flake surface. After the top hBN flake is picked up (Fig.~\ref{Fig1}p), we repeat this procedure by picking up subsequent layers of 2D materials, e.g. graphene (Fig.~\ref{Fig1}q), using the top hBN on mica to stack the heterostructure (Fig.~\ref{Fig1}i and r). Finally, the release mechanism relies on the stronger adhesion between the top and bottom hBN sheets in comparison to the mica-hBN. First, we pre-heat the substrate with bottom hBN to a higher temperature in a range between $120-180^{\circ}$ C (Fig.~\ref{Fig1}m). This is partly to decrease the adhesion between mica and hBN. The heterostructure is then brought in contact with the bottom hBN flake and the contact front line is moved across the heterostructure by varying the substrate temperature (Fig.~\ref{Fig1}s). It is important to leave a fraction of the bottom hBN surface not contacted by mica to avoid accidental pick up of the bottom hBN flake (Fig.~\ref{Fig1}n and t). This way, in contrast to existing polymer-based methods \cite{PDMSCastellanosGomez2014, PCZomer2014, PPCPizzocchero2016}, a selective release can be achieved by releasing only the stack of interest over the bottom hBN flake, leaving all other unneeded picked-up flakes on the stamp. A similar protocol is also applicable for the transfer using mica cantilevers. We demonstrate the typical heterostructure assembly process in detail in the Supplementary Video~1. 

\begin{figure}
\begin{center}
	\includegraphics[width=\linewidth]{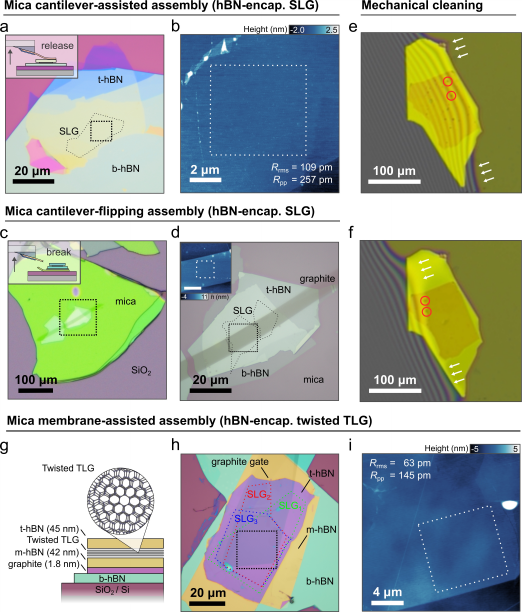}
	\caption{Examples of typical heterostructures assembled using mica-assisted methods. \textbf{a} Optical image of single-layer graphene (SLG) fully encapsulated in hBN via mica cantilever-assisted assembly; inset shows the standard release mechanism. \textbf{b} Contact mode atomic force microscopy (AFM) image of the same stack in the dashed black square in \textbf{a}, with surface roughness measured in the dashed white square. \textbf{c} Optical image of SLG encapsulated in hBN on a graphite gate electrode, released by breaking off the mica cantilever, which remains as the bottom layer. \textbf{d} Close up of the area indicated by the dashed black square in \textbf{c}. Inset: Contact mode-AFM image of the same stack scanned in the area indicated by the dashed black square in \textbf{d}. The root-mean-squared and peak-to-peak surface roughness in the dashed white square are $R_{\rm rms}=82$ pm and $R_{\rm pp}=201$ pm, respectively. Scale bar is $5$ $\mu$m. \textbf{e-f} Optical images showing mechanical cleaning using a mica membrane; white arrows indicate contact line movement, and red circles track contamination pockets being displaced. \textbf{g} Schematic of the multilayer heterostructure assembly, consisting of hBN-encapsulated twisted tri-layer graphene (TLG) and a graphite backgate electrode. \textbf{h} Optical image of the heterostructure with corresponding layers indicated. \textbf{i} Contact mode-AFM image of the dashed black square area in \textbf{d} with the surface roughness measured within the dashed white square. Coloured dashed lines indicate the outline of the relevant flake. t-hBN, m-hBN and b-hBN in \textbf{a}, \textbf{d} and \textbf{h} are the top, middle and bottom hBN flakes, respectively. Additional Raman spectroscopy and X-ray photoelectron spectroscopy (XPS) characterisation of hBN-encapsulated SLG stacks (see Supplementary Note 3 and 4) reveal high quality of graphene as well as no contamination of such heterostructures, neither with accidentally exfoliated mica layers nor with typical ions and dopants of mica (Li, K or Na).}
	\label{Fig2}
\end{center}
\end{figure}
\subsection*{Atomically flat and clean surface}
We further study heterostructures stacked and released using mica-based methods, using atomic force microscopy (AFM) to assess the quality of interfaces and surface roughness. In Fig.~\ref{Fig2} we demonstrate three hBN and graphene-based heterostructures assembled using mica cantilevers (Fig.~\ref{Fig2}a-d) and membrane approach (Fig.~\ref{Fig2}g-i). We show that in the case of cantilevers, we can release these heterostructures by both the standard method described above (Fig.~\ref{Fig2}a and b) and by flipping and purposefully breaking off the mica cantilever, utilising it as a substrate (Fig.~\ref{Fig2}c and d). The latter case is especially beneficial when the heterostructure is assembled bottom-up without relying on the additional bottom hBN flake for release. In Fig.~\ref{Fig2}b-d we demonstrate that in the working area of these stacks, we achieve the contamination-free interfaces with root-mean-square roughness $R_{\rm rms}$ in the order of $100$ pm over the areas indicated with the white dotted boxes. Such pristine interfaces are achieved by the fact that the contact front between mica and the sample can squeeze out contamination pockets and bubbles away from useful areas of the sample at elevated release temperatures due to weak bonding of mica (repelling) with organic contaminants \cite{RepellingLi2006,RepellingDonaldson2013}, as shown in Fig.~\ref{Fig2}e and f similarly to polymer-based cleaning methods \cite{CleaningPurdie2018}.

\begin{figure*}
\begin{center}
	\includegraphics[width=\linewidth]{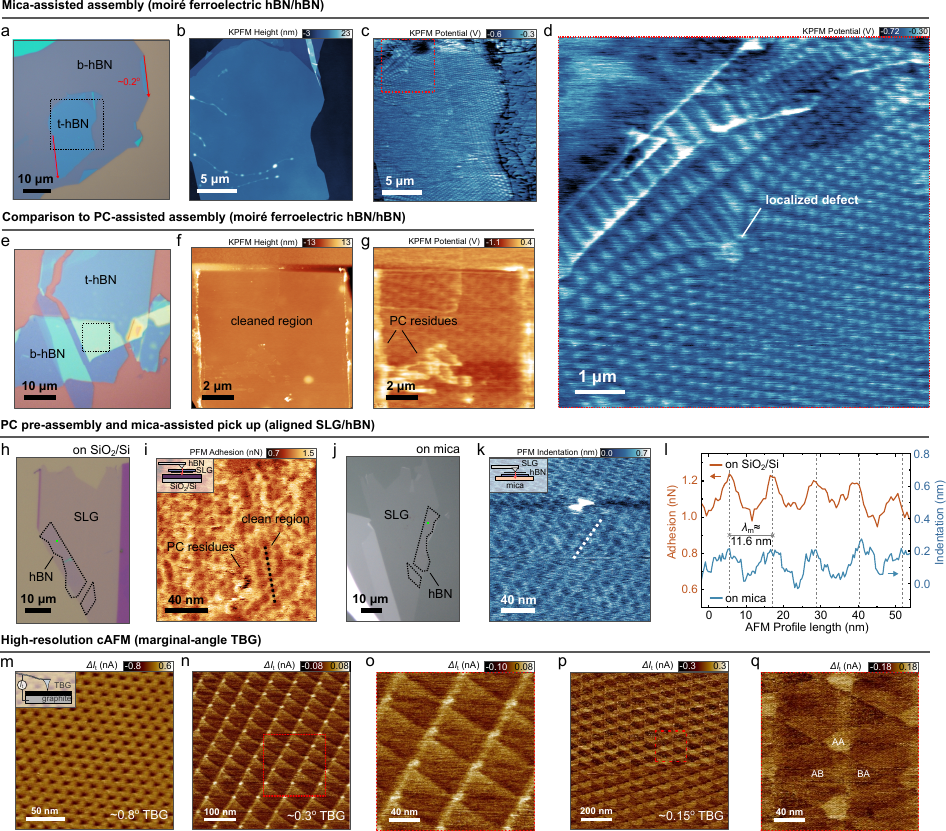}
	\caption{Examples of twisted layer heterostructures assembled using mica-assisted methods. \textbf{a} Optical image of marginally aligned ferroelectric hBN. \textbf{b} Height profile AFM image of the dashed black square area in \textbf{a}. \textbf{c} Kelvin probe force microscopy (KPFM) potential of the same AFM image in \textbf{b}. \textbf{d} Close-up KPFM scan of the dashed red square area in \textbf{c}. \textbf{e-g} Heterostructure of the same type assembled using the polycarbonate (PC) method for comparison. \textbf{e-g} follow the same structure as in \textbf{a-c}. We highlight that we are approaching a $50\%$ fabrication yield of such hBN-based moiré superlattice structures in \textbf{a-d}, limited only by the error in the determination of an even/odd number of layers of a particular flake. \textbf{h} Optical image of pre-assembled marginally aligned hBN on SLG using PC method. \textbf{i} Peak force microscopy (PFM) image of hBN/SLG moir\'{e} pattern in the area indicated with the green dot in \textbf{h}. \textbf{j} Optical image of the same stack in \textbf{h-i} picked up using mica-membrane. \textbf{k} PFM image of the moir\'{e} pattern in the area indicated with the green dot in \textbf{j}. \textbf{l} Comparison of moir\'{e} wavelength $\lambda_{\rm m}$ as measured on Si/SiO$_2$ along the dashed black line in \textbf{i} to that on mica membrane along the dashed white line in \textbf{k}. \textbf{m-q} High-resolution conductive atomic force microscopy (cAFM) tip-sample current variation $\Delta I_{\rm{t}}$ maps of twisted bilayer graphene (TBG) for different twist angles. \textbf{o} and \textbf{q} are close-up cAFM scans of the dashed red square area in \textbf{n} and \textbf{p}, respectively.}
	\label{Fig3}
\end{center}
\end{figure*}
\begin{figure*}
\begin{center}
	\includegraphics[width=\linewidth]{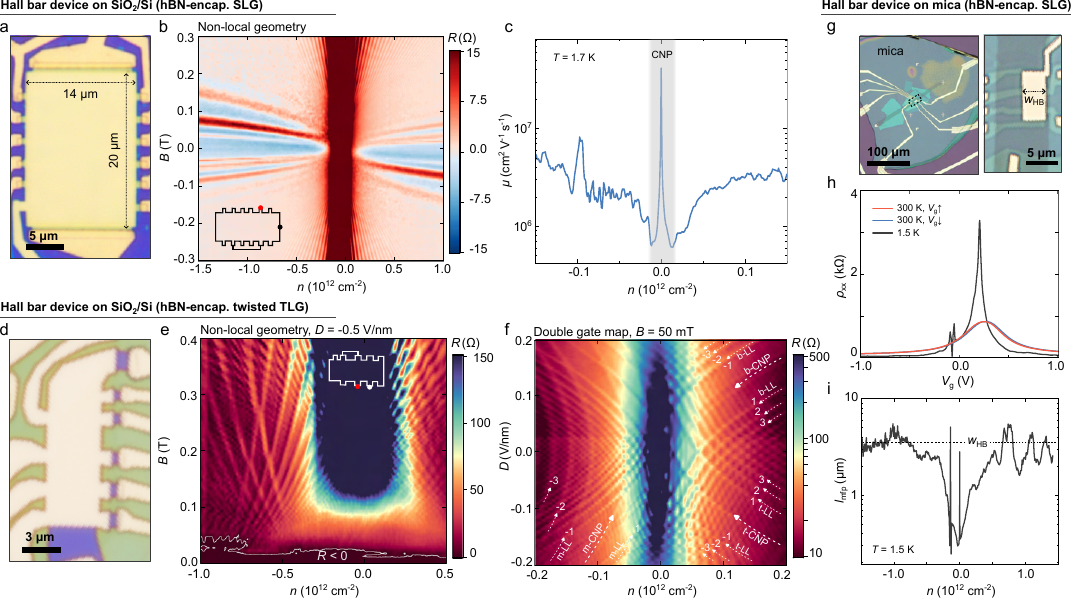}
	\caption{Low-temperature magneto-transport characterization of the heterostructures assembled using mica-assisted methods. \textbf{a} Optical image of a standard hBN-encapsulated SLG Hall bar device. \textbf{b} Non-local resistance measured as a function of charge carrier density $n$ and applied magnetic field $B$. Inset: Schematics of the measurement geometry of the sample in \textbf{a}. Red and black dots indicate source and drain, respectively. The black line indicates probe terminals. \textbf{c} Electron mobility $\mu$ as a function of $n$. The shaded grey area indicates the proximity of the charge neutrality point. \textbf{d} Optical image of the hBN-encapsulated twisted TLG Hall bar device made of heterostructure from Fig.~\ref{Fig2}\textbf{h-i}. \textbf{e} Non-local resistance map measured as a function of $n$ and $B$. Inset: Schematics of the measurement geometry of the sample in \textbf{d}. \textbf{f} Double-gate resistance map at $B=50$ mT. t-LL, m-LL and b-LL are Landau levels of the top, middle and bottom graphene layer, respectively. t-CNP, m-CNP and b-CNP are charge neutrality points in the top, middle and bottom graphene layer, respectively. \textbf{g} Optical images of the hBN-encapsulated SLG Hall bar device made of heterostructure form Fig.~\ref{Fig2}\textbf{c-d}. \textbf{h} Resistivity $\rho_{xx}$ of the device measured at $300$ and $1.5$ K. Red and blue curves are ascending and descending the gate voltage $V_{\rm g}$, respectively. \textbf{i} Mean-free path $l_{\rm mfp}$ of the same device. Dashed black line indicates the ballistic limit set by the device width $w_{\rm HB}$ from \textbf{g}.}
	\label{Fig4} 
\end{center}
\end{figure*}

Following the same recipe, we use the mica membrane-assisted approach to assemble more complex heterostructures with seven consecutive layers stacked into hBN/twisted trilayer graphene (TLG)/hBN/graphite/hBN, where each graphene is purposefully misaligned at high angles of $20$$^{\circ}$ (Fig.~\ref{Fig2}g-h). In Fig.~\ref{Fig2}i we demonstrate pristine interface quality over areas more than $100$ $\mu$m$^2$ with $R_{\rm rms}$ bellow $100$ pm (white dotted box). We emphasise that in all panels of Fig.~\ref{Fig2} stacks were characterised as-is following the release without any annealing, submerging in solvents or any other additional cleaning procedures. 

\subsection*{Fabrication and characterisation of moiré superlattices}
Polymer-free assembly yields an atomically flat top surface of the stack, making mica-assisted transfer a perfect tool for scanning probe microscopy (SPM) applications. In Fig.~\ref{Fig3}a, we assemble a marginally twisted hBN stack by picking up a piece of a cracked hBN flake using a mica membrane and releasing it on the same flake at a temperature of $50^{\circ}$C.  We further characterise the stack with Kelvin Probe Force Microscopy (KPFM) and reveal an array of triangular ferroelectric moiré patterns \cite{hBNMoore2021, hBNWoods2021,hBNYasuda2021} (Fig.~\ref{Fig3}b,c). The ferroelectric domains {of the moiré superlattice span across the whole top hBN flake ($\sim600$ $\mu$m$^2$). The ultimate surface and interlayer interface quality of the stack is further highlighted in Fig.~\ref{Fig3}d, where a single localized sub-micron size bubble defect between the hBN layers induces a continuous deformation of the ferroelectric domain pattern, opening up possibilities for effortless studies of strain effects and manipulation of 2D material superlattices \cite{Ding2024}. 

Furthermore, we demonstrate our results in contrast to KPFM studies of the moiré ferroelectric hBN/hBN structure assembled by standard PC-based methods (Fig.~\ref{Fig3}e). Here, after dissolving most of the PC in dichloromethane for several hours, polymer residues accumulated on the surface of the stack completely screen the moiré ferroelectric potential unless the surface is mechanically cleaned from residues using contact mode AFM (Fig.~\ref{Fig3}f and g). Another limitation of polymer-based methods that we address is layer misalignment \cite{Wang2016} due to the use of elevated temperatures to pick up and stack the flakes \cite{Martin2025}. In Fig.~\ref{Fig3}h, we show a pre-assembled marginally aligned single-layer graphene (SLG) on hBN. We characterise the aligned heterostructure with peak force microscopy (PFM), observing a clear hexagonal moiré superlattice of SLG/hBN between the traces of PC residues (Fig.~\ref{Fig3}i). Following that, we pick up the heterostructure using mica stamp at $50^{\circ}$C (Fig.~\ref{Fig3}j) and characterize the structure again with the same AFM cantilever using a mica membrane as a substrate (Fig.~\ref{Fig3}k). In Fig.~\ref{Fig3}l we show that picking up the marginally aligned SLG/hBN heterostructure on mica does not affect the moiré wavelength and thus the twist angle between the layers remains intact. This signifies the potential for mica-based assembly methods to not only stack moiré heterostructures but also to manipulate those pre-assembled and pre-characterised structures while maintaining the angle.

The capability of exposing conductive 2D layers in moiré heterostructures while maintaining their precisely aligned twist angle, together with pristine and residue-free surface without any post-treatment, is particularly attractive for conducting atomic force microscopy (cAFM) and related scanning tunnelling microscopy (STM) studies of twisted bilayers of graphene (TBG) \cite{TBGKerelsky2019,TBGHuang2018,TBGLiu2020}. In Fig~\ref{Fig3}m-q, we demonstrate a set of marginal-angle TBG devices fabricated by the mica-membrane method, with a sub-$1^{\circ}$ mismatch deterministically set in their twist angles to tune their moiré period and resulting local conducting properties. We place all TBG samples on top of thick graphite flakes, and characterise these using high-resolution cAFM (see Methods), under room-temperature conditions and a small bias applied between the sample and the tip. For the largest angle of $0.8^{\circ}$ in Fig~\ref{Fig3}m, we observed a highly uniform hexagonal moiré superlattice with $\sim17.1$ nm period. As we reduce the twist angle to $0.3^{\circ}$ (Fig~\ref{Fig3}n-o) and then $0.15^{\circ}$ (Fig~\ref{Fig3}p-q), the moiré period increases, revealing triangular AA, AB and BA domains in TBG as well as the domain wall structure between these \cite{TBGHuang2018}. These measurements demonstrate a highly uniform moiré superlattice with consistent period and clean surfaces across up to $1$ $\mu$m$^2$, allowing its detailed cAFM/STM studies on both the single moiré period level and the realistic optoelectronic device scales.}

\subsection*{Electronic quality of Hall bar devices}
Recent progress in device fabrication has enabled a significant reduction in charge disorder and inhomogeneity, bringing electron transport devices closer to their ultimate performance limits \cite{GenerationsRhodes2019, babich2025milli,geim2025proximity} and enabling the study of large-scale collective effects \cite{PhononOscillations2019}. Here we show that, in addition to providing ultraclean surfaces with simultaneous control of the twist angle, the mica-based method achieves graphene device quality at the level of the current state-of-the-art devices (see Supplementary Note 6). This includes devices with Coulomb screening that have already reached the ultimate performance limit, beyond which no further improvement is expected \cite{babich2025milli,geim2025proximity}.

Figure~\ref{Fig4}a shows a large-area hBN/SGL/hBN Hall-bar device fabricated on a SiO$_2$/Si substrate, with dimensions of $20 \times 14$ $\mu$m$^2$. The device was patterned in a bubble-free region of the heterostructure, highlighting the capability of our method to produce uniform, macroscopic interfaces between two-dimensional crystals. At liquid-helium temperatures, electron transport is limited by the device dimensions, indicating a mean free path comparable to the Hall-bar width. To demonstrate this, Fig.~\ref{Fig4}b presents non-local bend-resistance measurements in a multiterminal geometry (inset of Fig.~\ref{Fig4}b) under an applied magnetic field at $T=1.7$ K. At $B = 0$ T, a negative non-local resistance is observed, consistent with ballistic electron transport across the device width. With increasing magnetic field, rapidly decaying oscillations emerge, corresponding to magnetic electron focusing - another hallmark of ballistic transport. Consistently, the electron mobility $\mu$ at high carrier density $n$ reaches $4 \times 10^6$ cm$^2$V$^{-1}$s$^{-1}$, limited by the dimensions of the Hall-bar device (Fig.~\ref{Fig4}c). 

We further test the limits of our fabrication technique on a more complex heterostrucure in fabricating a Hall bar device from a 7-layer heterostructure based on a high-angle twisted trilayer graphene featured in Fig.~\ref{Fig2}i-g \cite{babich2025milli} (Fig.~\ref{Fig4}d). The graphene layers in such a structure are effectively decoupled by a momentum space mismatch, which results in the fan diagrams featuring Landau fans of all three layers (see Fig. \ref{Fig4}e) \cite{babich2025milli}. We, again, measure the device in a non-local bend geometry to probe both quantisation and ballisticity. In Fig. \ref{Fig4}e, we show a resistance map measured as a function of external magnetic field, $B$ and total charge carrier density, $n$, which features negative resistivity at zero magnetic field, indicating ballistic transport. Moreover, Landau fans exhibit a low onset of quantisation of 5$\pm$1 mT \cite{babich2025milli}, indicating a low disorder. In Fig.~\ref{Fig4}f we demonstrate a double gate map with more than $20$ distinct Landau levels observed at $B$ as small as $50$ mT. We note that we report magneto-transport effects in this device in detail in a separate study \cite{babich2025milli}. This puts devices fabricated by mica-assisted methods on the level comparable with the highest quality graphene Hall bar devices reported in the literature (see Supplementary Note 6) \cite{babich2025milli,geim2025proximity,Alexey2023, Manchester2023}.

\begin{figure*}
\begin{center}
	\includegraphics[width=\linewidth]{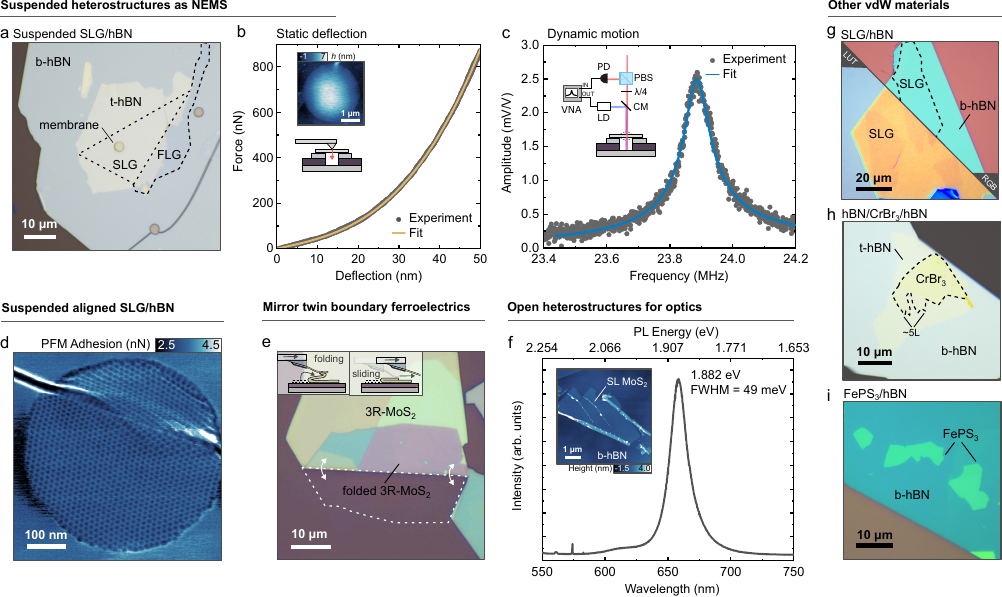}
	\caption{Further examples of mica-assembled functional devices. \textbf{a} Optical image of the suspended SLG/hBN heterostructure over the cavity in hBN/SiO$_2$ substrate. \textbf{b} Force-deflection AFM experiment on the membrane in \textbf{a}. Yellow line - fit to the standard continuum mechanics model \cite{EhBNFalin2017, EGrapheneLee2008} for Young's modulus $E=851\pm4$ GPa and pre-tension $N_0=1.3\pm0.2$ Nm$^{-1}$. Insets: the top panel is the AFM height map of the membrane. The bottom panel is the schematic of the experimental arrangement. \textbf{c} Dynamic measurement of the membrane's fundamental mechanical resonance mode. Blue line - fit to a linear damped harmonic oscillator model for resonance frequency $f_0=23.884\pm0.005$ MHz and quality factor $Q=314\pm3$. Inset: Schematic of the interferometric measurement setup. The detection scheme is described in detail in previous works \cite{Davidovikj2016, Siskins2020}. \textbf{d} PFM adhesion map of a suspended aligned SLG/hBN(1-2 layers) heterostructure. \textbf{e} Optical image of a folded ferroelectric 3R-MoS$_2$. Inset: schematic of folding (left) and release (right) procedures. \textbf{f} PL emission from a SL MoS$_2$ flake transferred onto hBN substrate measured at room temperature. Inset: AFM height map of the flake. Optical images of \textbf{g} large area uncapped SLG on hBN, \textbf{h} hBN-encapsulated few-layer CrBr$_3$, and \textbf{i} few-layer FePS$_3$ on hBN heterostructures.}
	\label{Fig5}
\end{center}
\end{figure*}
Furthermore, we fabricated a Hall bar device out of the structure featured in Fig.~\ref{Fig2}c and d, where the heterostructure was assembled on a freestanding mica cantilever and subsequently released by flipping and breaking it off. Here, we utilize the bottom graphite as a bottom gate and Ti/Au as a top gate, contacting the graphene using a standard fabrication workflow, guiding metal contact over the several hundred-nanometer-thick mica substrate (Fig.~\ref{Fig4}g). As we show in Fig.~\ref{Fig4}h, we observe neither hysteresis \cite{Mohrmann2014} nor large asymmetries of the Dirac peak \cite{Low2014} for ascending and descending gate voltage sweeps at $T=300$ K, which is a typical issue observed in graphene-on-mica devices. Likewise, cooling down the device to $T=1.5$ K shows rather standard behaviour of SLG Hall bars. Moreover, we observed that the mean free path $l_{\rm mfp}$ of electrons in our device is limited by the width of the Hall bar (Fig.\ref{Fig4} i), indicating the high quality of the device even when mica is kept in the stack as a substrate.

\subsection*{Suspending 2D material heterostructures as membranes}
Since mica membranes and cantilevers do not leave any polymeric residues on the surface, treatment by solvents or high-temperature annealing is not required after the stack is made. This advancement in fact holds a crucial potential for another field of study - nanomechanics of suspended 2D material membranes. Due to the poor yield of suspended membrane devices after contact with liquids, the fabrication of 2D material membrane-based nano-electromechanical systems (NEMS) from complex heterostructures practically is limited to top-down transfer using PDMS \cite{Siskins2022,PDMSCastellanosGomez2014} and transfer of pre-grown heterostructures with chemical vapour deposition \cite{Kim2018}, which produces wrinkles \cite{Davidovikj2016} and heavily contaminated interfaces \cite{Kim2018}. We demonstrate that the mica-assisted membrane transfer technique solves these issues. As shown in Fig.~\ref{Fig5}a, using the same approach and conditions as before, we suspend SGL/hBN heterostructure over the hole etched in the bottom hBN flake, creating a heterostructure membrane of $d=3$ $\mu$m in diameter. The inset of Fig.~\ref{Fig5}b shows the pristine surface of the resulting membrane as measured with PFM. 

To demonstrate the membrane behaviour of the suspended stack, we show the typical static force-deflection AFM experiment in Fig.~\ref{Fig5}b, well in agreement with previous reports \cite{EhBNFalin2017, EGrapheneLee2008}. We further demonstrate an example of the dynamic nanomechanical experiment in Fig.~\ref{Fig5}c (see Methods), showing the resonance peak of the fundamental vibrational mode of the membrane, which is well-described by the linear damped harmonic oscillator model in line with previous reports \cite{hBNresZheng2017,hBNres2020circular}. This fabrication principle can be extended further and applied to aligned graphene and hBN stacks. We demonstrate this by suspending the aligned SLG/hBN moir\'{e} superlattice over a pre-etched $d=0.5$ $\mu$m cavity in hBN. Fig.~\ref{Fig5}d shows PFM adhesion map of the suspended superlattice, revealing a hexagonal moir\'{e} pattern over the whole membrane surface. This demonstrates a successful fabrication of contamination-free suspended moir\'{e} superlattices as membranes that in future can be used for a variety of advanced nanomechanical experiments as well as electron transport and SPM devices, properties of which are unaffected by the substrate.

\subsection*{Heterostructures beyond graphene}
Despite the lower adhesion of other materials to mica in comparison to the mica-hBN pair, the same fabrication strategy can be applied to stack and manipulate 2D crystals beyond hBN. For instance, we were able to use mica cantilevers to pick up a part of large 3R-MoS$_2$ flakes, fold these onto themselves, and carefully delaminate and release these by sliding the flake off the cantilever, moving in parallel to the substrate surface \cite{AdvMatLee2025}. In Fig.~\ref{Fig5}e we demonstrate the resulting folded 3R-MoS$_2$ flake, which exhibits a low surface roughness of $R_{\rm{pp}}=244$ pm (folded region) and $R_{\rm{pp}}=182$ pm (unfolded region) together with high-quality field effect transistor performance and low hysteresis \cite{3RYang2023} (see Supplementary Note 7). Due to the ferroelectric nature of 3R-MoS$_2$, these folded structures with pristine interfaces hold the potential to host charge carriers at their mirror twin boundaries \cite{McHugh2024}, which up to date was challenging to observe experimentally due to limitations of standard fabrication methods.

We further demonstrate examples of stacks terminated with an open monolayer of MoS$_2$, SLG and few-layer FePS$_3$ transferred using the same procedure (Fig.~\ref{Fig5}f, g and i) as well as successfully encapsulated air-sensitive thin CrBr$_3$ crystals (Fig.~\ref{Fig5}h). In some cases, picking up the top 2D material other than hBN requires mild UV/O$_3$ treatment at 60$^{\circ}$ C for 12 minutes to improve the adhesive properties of mica. This also ensures that large areas of flakes as thin as a single monolayer stay intact and free of contamination, as shown in Fig.~\ref{Fig5}e. The resulting heterostructures, such as uncapped MoS$_2$ on hBN, show a surface roughness $R_{\rm rms}=136$ pm and optical performance with a bright photoluminescence signal at $1.882$ eV with full width at half maximum $\rm{FWHM}=49$ meV in air and at room temperature (Fig.~\ref{Fig5}f). We also note that this capability of exposing conductive 2D layers in heterostructures residue-free is particularly attractive for STM, cAFM and related studies (see Fig.~\ref{Fig3}m-q), as it enables sample fabrication without thorough post-treatment of the surface. All this demonstrates the versatility and broad applicability of our method, which allows the polymer-free fabrication of complex heterostructures for various applications, such as light-emitting diodes based on air- and temperature-sensitive materials \cite{Zawadzka2025}.

\section*{Discussion}
We have introduced and validated a polymer-free, deterministic fabrication method for 2D material heterostructures using muscovite crystal membranes, leveraging its natural van der Waals properties and temperature-dependent adhesion. This technique overcomes key limitations of existing methods, achieving large-area cleanliness, control over twist angles within the stack, and compatibility with a diverse range of materials and devices. A key strength of this approach lies in its compatibility with existing nanofabrication processes, as it requires no modifications to standard transfer setups, and can be readily adopted by simply replacing the polymer PC or PPC in commonly used PDMS/PC(PPC) stamps \cite{PPCPizzocchero2016,PPCWang2013,PCZomer2014}. At the same time, it is very cost-effective with a price for a single AFM-grade mica crystal around $3$ USD, making the cost of a single cantilever made of mica less than $10$ cents. 

However, like any fabrication method, it has certain limitations in the current state of development. At present, the mica method is best suited for high-performance, small-batch research devices, with typical lateral dimensions in the microscale (up to $\sim200$ $\mu$m), particularly excelling in assembling precisely aligned stacks and suspending membranes. One notable challenge to its scalability is the difficulty in reliably transferring these aligned heterostructures onto semiconductor industry-compatible substrates, like SiO$_2$/Si, without relying on a bottom 2D material flake, like hBN or graphite (which is compatible with the vast majority of research devices), or breaking off mica cantilevers, while maintaining the twist angle between the layers. However, we believe that these limitations are practical rather than fundamental. Natural mica exists in macroscopically large single crystals, which could enable large-area transfer in the future if compatible growth and release processes are developed. We also argue that for device geometries where the twist angle is not important, a sliding approach, similar to that in Fig.~\ref{Fig5}e and ref.~\cite{AdvMatLee2025}, can provide a reliable release mechanism, while retaining the benefits of clean, polymer-free surfaces and interfaces.

Another challenge lies in the reliance on the mechanical and adhesive properties of muscovite crystals, which may not be universally optimal for all 2D materials. For instance, materials with weaker van der Waals interactions or surfaces with low adhesion to mica may pose challenges during the pick-up or release stages. This becomes particularly important when a material has high sensitivity to environmental factors, like high humidity. Importantly, most of the heterostructures reported in this work were assembled reliably under typical laboratory conditions, without strict humidity control and additional optimisation (see Supplementary Note 8). However, operating in controlled environments, such as gloveboxes or under low-humidity conditions, can further help to mitigate these issues, including scenarios when high humidity levels alter the adhesion forces between mica and 2D materials (see Methods).

To summarize, our results establish the mica-assisted approach as a versatile and reliable method for the assembly of complex van der Waals heterostructures. Our stacks systematically demonstrate atomically clean, contamination-free, and flat surfaces and interfaces without any post-processing. This enables fabrication of highly uniform moir\'{e} superlattices as well as high-quality devices matching ultra-high mobility and performance of the best graphene-based Hall bar devices currently available. Beyond conventional stacks, the method enables a wide variety of device architectures, including suspended unencapsulated moir\'{e} superlattice membranes. Its broad material compatibility, which ranges from robust crystals (graphene, hBN, MoS$_2$) to air-sensitive (FePS$_3$, CrBr$_3$), further underscores its universality. We anticipate that this approach can also be extended to non-van der Waals materials, such as freestanding complex oxides \cite{Lu2016}, and to more exotic 2D systems, such as monolayer amorphous carbon \cite{Toh2020}, thereby broadening its impact beyond 2D materials.

\section*{Methods}
\textbf{Mica membrane and cantilever preparation.}
We cleave mica crystals using standard mechanical exfoliation methods \cite{Low2012}, obtaining a few hundred-nanometer-thin mica layers of lateral dimensions reaching up to $8\times 8$ mm supported by the adhesive tape. We developed two main transfer strategies: i) utilizing PDMS-supported mica membranes for classical 2D heterostructure assembly \cite{PCZomer2014, PPCPizzocchero2016} (Fig.~\ref{Fig1}a-d), and ii) as freestanding mica cantilevers for polymer-free fabrication protocols \cite{Manchester2023} (Fig.~\ref{Fig1}e-h). We typically use Nitto ELP BT-50E-FR or a few layers of Scotch Magic tape to support the membrane. In the first case, we suspend a mica membrane over a hole of $5-6$ mm diameter punched in an adhesive tape support, as shown in Fig.~\ref{Fig1}b. Then, we cover a rectangular PDMS piece placed on a microscope glass slide with the mica membrane and fix the adhesive tape on the sides of the glass slide (Fig.~\ref{Fig1}c). PDMS piece provides the required support for the membrane and tensions it to smooth out the folds and wrinkles. Likewise, in the second case, we cleave freestanding mica pieces of similar thickness using the edge of adhesive tape, creating transparent cantilevers of triangular shape (Fig.~\ref{Fig1}f). Then, these are transferred on the angled metal support ($\sim15^{\circ}$) and fixed on the microscope glass slide, as shown in Fig.~\ref{Fig1}g. The metal support ensures a reproducible contact angle between the cantilever and the substrate as well as a clear view during the subsequent flake pick-up and release (Fig.~\ref{Fig1}h).

The heterostructure assembly using the presented method can then be reliably handled both in the Argon glove box and air conditions, provided that the humidity level of the ambient atmosphere is lower than $\sim50\%$ to avoid parasitic water adsorption to mica \cite{Koishi2022, Xu2010}. Hence, we keep both source bulk mica crystals and prepared mica stamps/cantilevers under either an inert gas atmosphere (e.g. Ar) or in a dry vacuum desiccator before the transfer process.

\textbf{Conductive atomic force microscopy.}
The cAFM characterisation was carried out using a Bruker Dimension Icon AFM, equipped with a Nanoscope VII controller and a Tunnelling Current (PFTUNA) module, under standard laboratory conditions in ambient air and relative humidity below $56\%$. Scanning was performed in contact mode with CONTV-PT, Pt-coated Si tip (tip radius $\sim25$ nm, $k\approx0.2$ Nm$^{-1}$). The scan rate during the lateral scans was 0.5-1 Hz. Fig. 3m was acquired with a sample bias of 1 mV, a deflection setpoint of -0.1 V, and a resolution of 256 × 256; Fig. 3n with 0.4 mV, 0.02 V, and 512 × 512; Fig. 3o with 0.4 mV, 0.02 V, and 256 × 256; Fig. 3p with 7 mV, 0.08 V, and 512 × 512; and Fig. 3q with 7 mV, 0.05 V, and 256 × 256, respectively. A typical total measured current was in a range of 1-5 nA. Note that the current maps shown in Fig. 3m-q were processed using a first-order flattening procedure in NanoScope Analysis 3.00 software.

\textbf{Humidity effects on vdW material adhesion.}
To assess the role of humidity in our transfer protocols, we measured the adhesion forces of mica cantilevers to graphite, hBN and SiO$_2$ surfaces using AFM force-deflection technique \cite{Rokni2020} and a gold-plated mica cantilever (see Supplementary Note 5). We found that the adhesion force between mica and hBN increases significantly up to almost an order of magnitude when exposed to high humidity conditions ($>60\%$), which however is in line with reports of an increased adhesion of 2D materials to polymer transfer substrates due to water adsorption in conventional fabrication techniques \cite{Ma2017, Cai2022}. In contrast, under low humidity conditions, raising the temperature of a substrate surface well above the room temperature systematically reduces adhesion forces for all studied material pairs until the plateau is established around $80-100$ $^{\circ}$C. This observation suggests that working outside of the inert atmosphere of a glove box, elevated temperatures are crucial to minimise the sensitivity of the transfer protocol to the humidity of ambient air. We argue, however, that this requirement is not specific to our transfer technique only and may be equally applicable also to other methods of hBN-based heterostructure assembly \cite{Manchester2023, PCZomer2014, Ma2017, Cai2022}.

\textbf{Laser interferometry measurements.}
The sample is mounted on a $xy$ positioning stage inside an optical cryostat Attocube attoDRY800 connected to a laser interferometry setup. The membrane is cooled to the base temperature of $T=4$ K in high vacuum. We use a power-modulated diode blue laser ($\lambda=450$ nm) to excite the membrane to motion, and a HeNe red laser ($\lambda=633$ nm) to interferometrically read out the amplitude of its motion, which is further processed by a vector network analyser \cite{Siskins2020}. Laser spot size is on the order of $\sim1$ $\mu$m. All measurements were performed at incident laser powers of $<1$ $\mu$W.

\subsection*{Data availability}
Relevant data supporting the key findings of this study are available within the article and the Supplementary Information file. All raw data generated during the current study are available from the corresponding authors upon request.

\bibliographystyle{naturemag}

\begin{thebibliography}{10}
\expandafter\ifx\csname url\endcsname\relax
  \def\url#1{\texttt{#1}}\fi
\expandafter\ifx\csname urlprefix\endcsname\relax\def\urlprefix{URL }\fi
\providecommand{\bibinfo}[2]{#2}
\providecommand{\eprint}[2][]{\url{#2}}

\bibitem{HeteroGeim2013}
\bibinfo{author}{Geim, A.~K.} \& \bibinfo{author}{Grigorieva, I.~V.}
\newblock \bibinfo{title}{Van der {W}aals heterostructures}.
\newblock \emph{\bibinfo{journal}{Nature}} \textbf{\bibinfo{volume}{499}},
  \bibinfo{pages}{419–425} (\bibinfo{year}{2013}).

\bibitem{HeteroNovoselov2016}
\bibinfo{author}{Novoselov, K.~S.}, \bibinfo{author}{Mishchenko, A.},
  \bibinfo{author}{Carvalho, A.} \& \bibinfo{author}{Castro~Neto, A.~H.}
\newblock \bibinfo{title}{{2D} materials and van der {W}aals heterostructures}.
\newblock \emph{\bibinfo{journal}{Science}} \textbf{\bibinfo{volume}{353}},
  \bibinfo{pages}{aac9439} (\bibinfo{year}{2016}).

\bibitem{HeteroLiang2019}
\bibinfo{author}{Liang, S.}, \bibinfo{author}{Cheng, B.}, \bibinfo{author}{Cui,
  X.} \& \bibinfo{author}{Miao, F.}
\newblock \bibinfo{title}{Van der {W}aals heterostructures for
  high‐performance device applications: Challenges and opportunities}.
\newblock \emph{\bibinfo{journal}{Adv. Mater.}} \textbf{\bibinfo{volume}{32}},
  \bibinfo{pages}{1903800} (\bibinfo{year}{2019}).

\bibitem{ReviewFrisenda2018}
\bibinfo{author}{Frisenda, R.} \emph{et~al.}
\newblock \bibinfo{title}{Recent progress in the assembly of nanodevices and
  van der {W}aals heterostructures by deterministic placement of {2D}
  materials}.
\newblock \emph{\bibinfo{journal}{Chem. Soc. Rev.}}
  \textbf{\bibinfo{volume}{47}}, \bibinfo{pages}{53–68}
  (\bibinfo{year}{2018}).

\bibitem{ReviewPham2024}
\bibinfo{author}{Pham, P.~V.} \emph{et~al.}
\newblock \bibinfo{title}{Transfer of {2D} films: From imperfection to
  perfection}.
\newblock \emph{\bibinfo{journal}{ACS Nano}} \textbf{\bibinfo{volume}{18}},
  \bibinfo{pages}{14841–14876} (\bibinfo{year}{2024}).

\bibitem{Martin2025}
\bibinfo{author}{Díez-Mérida, J.} \emph{et~al.}
\newblock \bibinfo{title}{High-yield fabrication of bubble-free magic-angle
  twisted bilayer graphene devices with high twist-angle homogeneity}.
\newblock \emph{\bibinfo{journal}{Newton}} \textbf{\bibinfo{volume}{1}},
  \bibinfo{pages}{100007} (\bibinfo{year}{2025}).

\bibitem{Cao2018}
\bibinfo{author}{Cao, Y.} \emph{et~al.}
\newblock \bibinfo{title}{Correlated insulator behaviour at half-filling in
  magic-angle graphene superlattices}.
\newblock \emph{\bibinfo{journal}{Nature}} \textbf{\bibinfo{volume}{556}},
  \bibinfo{pages}{80–84} (\bibinfo{year}{2018}).

\bibitem{Cao2018_2}
\bibinfo{author}{Cao, Y.} \emph{et~al.}
\newblock \bibinfo{title}{Unconventional superconductivity in magic-angle
  graphene superlattices}.
\newblock \emph{\bibinfo{journal}{Nature}} \textbf{\bibinfo{volume}{556}},
  \bibinfo{pages}{43–50} (\bibinfo{year}{2018}).

\bibitem{Lu2019}
\bibinfo{author}{Lu, X.} \emph{et~al.}
\newblock \bibinfo{title}{Superconductors, orbital magnets and correlated
  states in magic-angle bilayer graphene}.
\newblock \emph{\bibinfo{journal}{Nature}} \textbf{\bibinfo{volume}{574}},
  \bibinfo{pages}{653–657} (\bibinfo{year}{2019}).

\bibitem{Guo2025}
\bibinfo{author}{Guo, Y.} \emph{et~al.}
\newblock \bibinfo{title}{Superconductivity in $5.0^{\circ}$ twisted-bilayer
  {WSe$_2$}}.
\newblock \emph{\bibinfo{journal}{Nature}} \textbf{\bibinfo{volume}{637}},
  \bibinfo{pages}{839–845} (\bibinfo{year}{2025}).

\bibitem{Zondiner2020}
\bibinfo{author}{Zondiner, U.} \emph{et~al.}
\newblock \bibinfo{title}{Cascade of phase transitions and {D}irac revivals in
  magic-angle graphene}.
\newblock \emph{\bibinfo{journal}{Nature}} \textbf{\bibinfo{volume}{582}},
  \bibinfo{pages}{203–208} (\bibinfo{year}{2020}).

\bibitem{WetReina2008}
\bibinfo{author}{Reina, A.} \emph{et~al.}
\newblock \bibinfo{title}{Transferring and identification of single- and
  few-layer graphene on arbitrary substrates}.
\newblock \emph{\bibinfo{journal}{J. Phys. Chem. C.}}
  \textbf{\bibinfo{volume}{112}}, \bibinfo{pages}{17741–17744}
  (\bibinfo{year}{2008}).

\bibitem{Fan2020}
\bibinfo{author}{Fan, S.}, \bibinfo{author}{Vu, Q.~A.}, \bibinfo{author}{Tran,
  M.~D.}, \bibinfo{author}{Adhikari, S.} \& \bibinfo{author}{Lee, Y.~H.}
\newblock \bibinfo{title}{Transfer assembly for two-dimensional van der {W}aals
  heterostructures}.
\newblock \emph{\bibinfo{journal}{2D Mater.}} \textbf{\bibinfo{volume}{7}},
  \bibinfo{pages}{022005} (\bibinfo{year}{2020}).

\bibitem{PMMAReina2009}
\bibinfo{author}{Reina, A.} \emph{et~al.}
\newblock \bibinfo{title}{Large area, few-layer graphene films on arbitrary
  substrates by chemical vapor deposition}.
\newblock \emph{\bibinfo{journal}{Nano Lett.}} \textbf{\bibinfo{volume}{9}},
  \bibinfo{pages}{30–35} (\bibinfo{year}{2009}).

\bibitem{PMMADean2010}
\bibinfo{author}{Dean, C.~R.} \emph{et~al.}
\newblock \bibinfo{title}{Boron nitride substrates for high-quality graphene
  electronics}.
\newblock \emph{\bibinfo{journal}{Nat. Nanotechnol.}}
  \textbf{\bibinfo{volume}{5}}, \bibinfo{pages}{722–726}
  (\bibinfo{year}{2010}).

\bibitem{PDMSCastellanosGomez2014}
\bibinfo{author}{Castellanos-Gomez, A.} \emph{et~al.}
\newblock \bibinfo{title}{Deterministic transfer of two-dimensional materials
  by all-dry viscoelastic stamping}.
\newblock \emph{\bibinfo{journal}{2D Mater.}} \textbf{\bibinfo{volume}{1}},
  \bibinfo{pages}{011002} (\bibinfo{year}{2014}).

\bibitem{PDMSYan2025}
\bibinfo{author}{Yan, H.} \emph{et~al.}
\newblock \bibinfo{title}{Clean transfer of two-dimensional materials using
  {UV}-ozone treated polydimethylsiloxane}.
\newblock \emph{\bibinfo{journal}{2D Mater.}} \textbf{\bibinfo{volume}{12}},
  \bibinfo{pages}{045011} (\bibinfo{year}{2025}).

\bibitem{PCZomer2014}
\bibinfo{author}{Zomer, P.~J.}, \bibinfo{author}{Guimarães, M. H.~D.},
  \bibinfo{author}{Brant, J.~C.}, \bibinfo{author}{Tombros, N.} \&
  \bibinfo{author}{van Wees, B.~J.}
\newblock \bibinfo{title}{Fast pick up technique for high quality
  heterostructures of bilayer graphene and hexagonal boron nitride}.
\newblock \emph{\bibinfo{journal}{Appl. Phys. Lett.}}
  \textbf{\bibinfo{volume}{105}}, \bibinfo{pages}{013101}
  (\bibinfo{year}{2014}).

\bibitem{PPCWang2013}
\bibinfo{author}{Wang, L.} \emph{et~al.}
\newblock \bibinfo{title}{One-dimensional electrical contact to a
  two-dimensional material}.
\newblock \emph{\bibinfo{journal}{Science}} \textbf{\bibinfo{volume}{342}},
  \bibinfo{pages}{614–617} (\bibinfo{year}{2013}).

\bibitem{PPCPizzocchero2016}
\bibinfo{author}{Pizzocchero, F.} \emph{et~al.}
\newblock \bibinfo{title}{The hot pick-up technique for batch assembly of van
  der {W}aals heterostructures}.
\newblock \emph{\bibinfo{journal}{Nat. Commun.}} \textbf{\bibinfo{volume}{7}},
  \bibinfo{pages}{11894} (\bibinfo{year}{2016}).

\bibitem{CleaningPurdie2018}
\bibinfo{author}{Purdie, D.~G.} \emph{et~al.}
\newblock \bibinfo{title}{Cleaning interfaces in layered materials
  heterostructures}.
\newblock \emph{\bibinfo{journal}{Nat. Commun.}} \textbf{\bibinfo{volume}{9}},
  \bibinfo{pages}{5387} (\bibinfo{year}{2018}).

\bibitem{BubbleIwasaki2020}
\bibinfo{author}{Iwasaki, T.} \emph{et~al.}
\newblock \bibinfo{title}{Bubble-free transfer technique for high-quality
  graphene/hexagonal boron nitride van der {W}aals heterostructures}.
\newblock \emph{\bibinfo{journal}{{ACS} Appl. Mater. {\&} Interfaces}}
  \textbf{\bibinfo{volume}{12}}, \bibinfo{pages}{8533–8538}
  (\bibinfo{year}{2020}).

\bibitem{Annett2016}
\bibinfo{author}{Annett, J.} \& \bibinfo{author}{Cross, G. L.~W.}
\newblock \bibinfo{title}{Self-assembly of graphene ribbons by spontaneous
  self-tearing and peeling from a substrate}.
\newblock \emph{\bibinfo{journal}{Nature}} \textbf{\bibinfo{volume}{535}},
  \bibinfo{pages}{271–275} (\bibinfo{year}{2016}).

\bibitem{Wang2016}
\bibinfo{author}{Wang, D.} \emph{et~al.}
\newblock \bibinfo{title}{Thermally induced graphene rotation on hexagonal
  boron nitride}.
\newblock \emph{\bibinfo{journal}{Phys. Rev. Lett.}}
  \textbf{\bibinfo{volume}{116}}, \bibinfo{pages}{126101}
  (\bibinfo{year}{2016}).

\bibitem{Tran2024}
\bibinfo{author}{Tran, S.~J.} \emph{et~al.}
\newblock \bibinfo{title}{Quantitative determination of twist angle and strain
  in van der {W}aals moiré superlattices}.
\newblock \emph{\bibinfo{journal}{Appl. Phys. Lett.}}
  \textbf{\bibinfo{volume}{125}}, \bibinfo{pages}{113106}
  (\bibinfo{year}{2024}).

\bibitem{CantileverJin2023}
\bibinfo{author}{Jin, K.} \emph{et~al.}
\newblock \bibinfo{title}{Assembly of arbitrary designer heterostructures with
  atomically clean interfaces}.
\newblock \emph{\bibinfo{journal}{Adv. Mater. Interfaces}}
  \textbf{\bibinfo{volume}{11}}, \bibinfo{pages}{2300658}
  (\bibinfo{year}{2023}).

\bibitem{AdvMatLee2025}
\bibinfo{author}{Lee, M.} \emph{et~al.}
\newblock \bibinfo{title}{Residue‐free fabrication of {2D} materials using
  van der {W}aals interactions}.
\newblock \emph{\bibinfo{journal}{Adv. Mater.}} \textbf{\bibinfo{volume}{37}},
  \bibinfo{pages}{2418669} (\bibinfo{year}{2025}).

\bibitem{Manchester2023}
\bibinfo{author}{Wang, W.} \emph{et~al.}
\newblock \bibinfo{title}{Clean assembly of van der {W}aals heterostructures
  using silicon nitride membranes}.
\newblock \emph{\bibinfo{journal}{Nat. Electron.}}
  \textbf{\bibinfo{volume}{6}}, \bibinfo{pages}{981–990}
  (\bibinfo{year}{2023}).

\bibitem{dePoel2014}
\bibinfo{author}{de~Poel, W.} \emph{et~al.}
\newblock \bibinfo{title}{Muscovite mica: Flatter than a pancake}.
\newblock \emph{\bibinfo{journal}{Surf. Sci.}} \textbf{\bibinfo{volume}{619}},
  \bibinfo{pages}{19–24} (\bibinfo{year}{2014}).

\bibitem{Sanchez2018}
\bibinfo{author}{Sanchez, D.~A.} \emph{et~al.}
\newblock \bibinfo{title}{Mechanics of spontaneously formed nanoblisters
  trapped by transferred {2D} crystals}.
\newblock \emph{\bibinfo{journal}{Proc. Natl. Acad. Sci. U.S.A.}}
  \textbf{\bibinfo{volume}{115}}, \bibinfo{pages}{7884–7889}
  (\bibinfo{year}{2018}).

\bibitem{MicaCastellanosGomez2012}
\bibinfo{author}{Castellanos-Gomez, A.} \emph{et~al.}
\newblock \bibinfo{title}{Mechanical properties of freely suspended atomically
  thin dielectric layers of mica}.
\newblock \emph{\bibinfo{journal}{Nano Res.}} \textbf{\bibinfo{volume}{5}},
  \bibinfo{pages}{550–557} (\bibinfo{year}{2012}).

\bibitem{RepellingLi2006}
\bibinfo{author}{Li, F.} \emph{et~al.}
\newblock \bibinfo{title}{Differences between tethered polyelectrolyte chains
  on bare mica and hydrophobically modified mica}.
\newblock \emph{\bibinfo{journal}{Langmuir}} \textbf{\bibinfo{volume}{22}},
  \bibinfo{pages}{4084–4091} (\bibinfo{year}{2006}).

\bibitem{RepellingDonaldson2013}
\bibinfo{author}{Donaldson, S.~H.} \emph{et~al.}
\newblock \bibinfo{title}{Asymmetric electrostatic and
  hydrophobic–hydrophilic interaction forces between mica surfaces and
  silicone polymer thin films}.
\newblock \emph{\bibinfo{journal}{ACS Nano}} \textbf{\bibinfo{volume}{7}},
  \bibinfo{pages}{10094–10104} (\bibinfo{year}{2013}).

\bibitem{hBNMoore2021}
\bibinfo{author}{Moore, S.~L.} \emph{et~al.}
\newblock \bibinfo{title}{Nanoscale lattice dynamics in hexagonal boron nitride
  moiré superlattices}.
\newblock \emph{\bibinfo{journal}{Nat. Commun.}} \textbf{\bibinfo{volume}{12}},
  \bibinfo{pages}{5741} (\bibinfo{year}{2021}).

\bibitem{hBNWoods2021}
\bibinfo{author}{Woods, C.~R.} \emph{et~al.}
\newblock \bibinfo{title}{Charge-polarized interfacial superlattices in
  marginally twisted hexagonal boron nitride}.
\newblock \emph{\bibinfo{journal}{Nat. Commun.}} \textbf{\bibinfo{volume}{12}},
  \bibinfo{pages}{347} (\bibinfo{year}{2021}).

\bibitem{hBNYasuda2021}
\bibinfo{author}{Yasuda, K.}, \bibinfo{author}{Wang, X.},
  \bibinfo{author}{Watanabe, K.}, \bibinfo{author}{Taniguchi, T.} \&
  \bibinfo{author}{Jarillo-Herrero, P.}
\newblock \bibinfo{title}{Stacking-engineered ferroelectricity in bilayer boron
  nitride}.
\newblock \emph{\bibinfo{journal}{Science}} \textbf{\bibinfo{volume}{372}},
  \bibinfo{pages}{1458–1462} (\bibinfo{year}{2021}).

\bibitem{Ding2024}
\bibinfo{author}{Ding, P.} \emph{et~al.}
\newblock \bibinfo{title}{Manipulation of moiré superlattice in twisted
  monolayer-multilayer graphene by moving nanobubbles}.
\newblock \emph{\bibinfo{journal}{Nano Lett.}} \textbf{\bibinfo{volume}{24}},
  \bibinfo{pages}{8208–8215} (\bibinfo{year}{2024}).

\bibitem{TBGKerelsky2019}
\bibinfo{author}{Kerelsky, A.} \emph{et~al.}
\newblock \bibinfo{title}{Maximized electron interactions at the magic angle in
  twisted bilayer graphene}.
\newblock \emph{\bibinfo{journal}{Nature}} \textbf{\bibinfo{volume}{572}},
  \bibinfo{pages}{95–100} (\bibinfo{year}{2019}).

\bibitem{TBGHuang2018}
\bibinfo{author}{Huang, S.} \emph{et~al.}
\newblock \bibinfo{title}{Topologically protected helical states in minimally
  twisted bilayer graphene}.
\newblock \emph{\bibinfo{journal}{Phys. Rev. Lett.}}
  \textbf{\bibinfo{volume}{121}}, \bibinfo{pages}{037702}
  (\bibinfo{year}{2018}).

\bibitem{TBGLiu2020}
\bibinfo{author}{Liu, Y.-W.} \emph{et~al.}
\newblock \bibinfo{title}{Tunable lattice reconstruction, triangular network of
  chiral one-dimensional states, and bandwidth of flat bands in magic angle
  twisted bilayer graphene}.
\newblock \emph{\bibinfo{journal}{Phys. Rev. Lett.}}
  \textbf{\bibinfo{volume}{125}}, \bibinfo{pages}{236102}
  (\bibinfo{year}{2020}).

\bibitem{GenerationsRhodes2019}
\bibinfo{author}{Rhodes, D.}, \bibinfo{author}{Chae, S.~H.},
  \bibinfo{author}{Ribeiro-Palau, R.} \& \bibinfo{author}{Hone, J.}
\newblock \bibinfo{title}{Disorder in van der {W}aals heterostructures of {2D}
  materials}.
\newblock \emph{\bibinfo{journal}{Nat. Mater.}} \textbf{\bibinfo{volume}{18}},
  \bibinfo{pages}{541–549} (\bibinfo{year}{2019}).

\bibitem{babich2025milli}
\bibinfo{author}{Babich, I.} \emph{et~al.}
\newblock \bibinfo{title}{Milli-{T}esla quantization enabled by tuneable
  {C}oulomb screening in large-angle twisted graphene}.
\newblock \emph{\bibinfo{journal}{Nat. Commun.}} \textbf{\bibinfo{volume}{16}},
  \bibinfo{pages}{7389} (\bibinfo{year}{2025}).

\bibitem{geim2025proximity}
\bibinfo{author}{Domaretskiy, D.} \emph{et~al.}
\newblock \bibinfo{title}{Proximity screening greatly enhances electronic
  quality of graphene}.
\newblock \emph{\bibinfo{journal}{Nature}} \textbf{\bibinfo{volume}{644}},
  \bibinfo{pages}{646–651} (\bibinfo{year}{2025}).

\bibitem{PhononOscillations2019}
\bibinfo{author}{Kumaravadivel, P.} \emph{et~al.}
\newblock \bibinfo{title}{Strong magnetophonon oscillations in extra-large
  graphene}.
\newblock \emph{\bibinfo{journal}{Nat. Commun.}} \textbf{\bibinfo{volume}{10}},
  \bibinfo{pages}{3334} (\bibinfo{year}{2019}).

\bibitem{Alexey2023}
\bibinfo{author}{Xin, N.} \emph{et~al.}
\newblock \bibinfo{title}{Giant magnetoresistance of {D}irac plasma in
  high-mobility graphene}.
\newblock \emph{\bibinfo{journal}{Nature}} \textbf{\bibinfo{volume}{616}},
  \bibinfo{pages}{270–274} (\bibinfo{year}{2023}).

\bibitem{EhBNFalin2017}
\bibinfo{author}{Falin, A.} \emph{et~al.}
\newblock \bibinfo{title}{Mechanical properties of atomically thin boron
  nitride and the role of interlayer interactions}.
\newblock \emph{\bibinfo{journal}{Nat. Commun.}} \textbf{\bibinfo{volume}{8}},
  \bibinfo{pages}{15815} (\bibinfo{year}{2017}).

\bibitem{EGrapheneLee2008}
\bibinfo{author}{Lee, C.}, \bibinfo{author}{Wei, X.}, \bibinfo{author}{Kysar,
  J.~W.} \& \bibinfo{author}{Hone, J.}
\newblock \bibinfo{title}{Measurement of the elastic properties and intrinsic
  strength of monolayer graphene}.
\newblock \emph{\bibinfo{journal}{Science}} \textbf{\bibinfo{volume}{321}},
  \bibinfo{pages}{385–388} (\bibinfo{year}{2008}).

\bibitem{Davidovikj2016}
\bibinfo{author}{Davidovikj, D.} \emph{et~al.}
\newblock \bibinfo{title}{Visualizing the motion of graphene nanodrums}.
\newblock \emph{\bibinfo{journal}{Nano Lett.}} \textbf{\bibinfo{volume}{16}},
  \bibinfo{pages}{2768–2773} (\bibinfo{year}{2016}).

\bibitem{Siskins2020}
\bibinfo{author}{Šiškins, M.} \emph{et~al.}
\newblock \bibinfo{title}{Magnetic and electronic phase transitions probed by
  nanomechanical resonators}.
\newblock \emph{\bibinfo{journal}{Nat. Commun.}} \textbf{\bibinfo{volume}{11}},
  \bibinfo{pages}{2698} (\bibinfo{year}{2020}).

\bibitem{Mohrmann2014}
\bibinfo{author}{Mohrmann, J.}, \bibinfo{author}{Watanabe, K.},
  \bibinfo{author}{Taniguchi, T.} \& \bibinfo{author}{Danneau, R.}
\newblock \bibinfo{title}{Persistent hysteresis in graphene-mica van der
  {W}aals heterostructures}.
\newblock \emph{\bibinfo{journal}{Nanotechnology}}
  \textbf{\bibinfo{volume}{26}}, \bibinfo{pages}{015202}
  (\bibinfo{year}{2014}).

\bibitem{Low2014}
\bibinfo{author}{Low, C.~G.}, \bibinfo{author}{Zhang, Q.},
  \bibinfo{author}{Hao, Y.} \& \bibinfo{author}{Ruoff, R.~S.}
\newblock \bibinfo{title}{Graphene field effect transistors with mica as gate
  dielectric layers}.
\newblock \emph{\bibinfo{journal}{Small}} \textbf{\bibinfo{volume}{10}},
  \bibinfo{pages}{4213–4218} (\bibinfo{year}{2014}).

\bibitem{Siskins2022}
\bibinfo{author}{Šiškins, M.} \emph{et~al.}
\newblock \bibinfo{title}{Nanomechanical probing and strain tuning of the
  {C}urie temperature in suspended {Cr$_2$Ge$_2$Te$_6$}-based
  heterostructures}.
\newblock \emph{\bibinfo{journal}{npj 2D Mater. Appl.}}
  \textbf{\bibinfo{volume}{6}}, \bibinfo{pages}{41} (\bibinfo{year}{2022}).

\bibitem{Kim2018}
\bibinfo{author}{Kim, S.}, \bibinfo{author}{Yu, J.} \& \bibinfo{author}{van~der
  Zande, A.~M.}
\newblock \bibinfo{title}{Nano-electromechanical drumhead resonators from
  two-dimensional material bimorphs}.
\newblock \emph{\bibinfo{journal}{Nano Lett.}} \textbf{\bibinfo{volume}{18}},
  \bibinfo{pages}{6686–6695} (\bibinfo{year}{2018}).

\bibitem{hBNresZheng2017}
\bibinfo{author}{Zheng, X.-Q.}, \bibinfo{author}{Lee, J.} \&
  \bibinfo{author}{Feng, P. X.-L.}
\newblock \bibinfo{title}{Hexagonal boron nitride nanomechanical resonators
  with spatially visualized motion}.
\newblock \emph{\bibinfo{journal}{Microsyst. Nanoeng.}}
  \textbf{\bibinfo{volume}{3}}, \bibinfo{pages}{17038} (\bibinfo{year}{2017}).

\bibitem{hBNres2020circular}
\bibinfo{author}{Kumar, R.} \emph{et~al.}
\newblock \bibinfo{title}{Circular electromechanical resonators based on
  hexagonal-boron nitride-graphene heterostructures}.
\newblock \emph{\bibinfo{journal}{Appl. Phys. Lett.}}
  \textbf{\bibinfo{volume}{117}}, \bibinfo{pages}{183103}
  (\bibinfo{year}{2020}).

\bibitem{3RYang2023}
\bibinfo{author}{Yang, T.~H.} \emph{et~al.}
\newblock \bibinfo{title}{Ferroelectric transistors based on
  shear-transformation-mediated rhombohedral-stacked molybdenum disulfide}.
\newblock \emph{\bibinfo{journal}{Nat. Electron.}}
  \textbf{\bibinfo{volume}{7}}, \bibinfo{pages}{29–38}
  (\bibinfo{year}{2023}).

\bibitem{McHugh2024}
\bibinfo{author}{McHugh, J.~G.}, \bibinfo{author}{Li, X.},
  \bibinfo{author}{Soltero, I.} \& \bibinfo{author}{Fal’ko, V.~I.}
\newblock \bibinfo{title}{Two-dimensional electrons at mirror and twistronic
  twin boundaries in van der {W}aals ferroelectrics}.
\newblock \emph{\bibinfo{journal}{Nat. Commun.}} \textbf{\bibinfo{volume}{15}},
  \bibinfo{pages}{6838} (\bibinfo{year}{2024}).

\bibitem{Zawadzka2025}
\bibinfo{author}{Zawadzka, N.} \emph{et~al.}
\newblock \bibinfo{title}{Electrically modulated light-emitting device driven
  by resonant and antiresonant tunneling between {Cr$_2$Ge$_2$Te$_6$}
  electrodes}.
\newblock \emph{\bibinfo{journal}{2D Mater.}} \textbf{\bibinfo{volume}{13}},
  \bibinfo{pages}{011001} (\bibinfo{year}{2025}).

\bibitem{Lu2016}
\bibinfo{author}{Lu, D.} \emph{et~al.}
\newblock \bibinfo{title}{Synthesis of freestanding single-crystal perovskite
  films and heterostructures by etching of sacrificial water-soluble layers}.
\newblock \emph{\bibinfo{journal}{Nat. Mater.}} \textbf{\bibinfo{volume}{15}},
  \bibinfo{pages}{1255–1260} (\bibinfo{year}{2016}).

\bibitem{Toh2020}
\bibinfo{author}{Toh, C.-T.} \emph{et~al.}
\newblock \bibinfo{title}{Synthesis and properties of free-standing monolayer
  amorphous carbon}.
\newblock \emph{\bibinfo{journal}{Nature}} \textbf{\bibinfo{volume}{577}},
  \bibinfo{pages}{199–203} (\bibinfo{year}{2020}).

\bibitem{Low2012}
\bibinfo{author}{Low, C.~G.} \& \bibinfo{author}{Zhang, Q.}
\newblock \bibinfo{title}{Ultra‐thin and flat mica as gate dielectric
  layers}.
\newblock \emph{\bibinfo{journal}{Small}} \textbf{\bibinfo{volume}{8}},
  \bibinfo{pages}{2178–2183} (\bibinfo{year}{2012}).

\bibitem{Koishi2022}
\bibinfo{author}{Koishi, A.}, \bibinfo{author}{Lee, S.~S.},
  \bibinfo{author}{Fenter, P.}, \bibinfo{author}{Fernandez-Martinez, A.} \&
  \bibinfo{author}{Bourg, I.~C.}
\newblock \bibinfo{title}{Water adsorption on mica surfaces with hydrophilicity
  tuned by counterion types ({Na}, {K}, and {Cs}) and structural fluorination}.
\newblock \emph{\bibinfo{journal}{J. Phys. Chem. C}}
  \textbf{\bibinfo{volume}{126}}, \bibinfo{pages}{16447–16460}
  (\bibinfo{year}{2022}).

\bibitem{Xu2010}
\bibinfo{author}{Xu, K.}, \bibinfo{author}{Cao, P.} \& \bibinfo{author}{Heath,
  J.~R.}
\newblock \bibinfo{title}{Graphene visualizes the first water adlayers on mica
  at ambient conditions}.
\newblock \emph{\bibinfo{journal}{Science}} \textbf{\bibinfo{volume}{329}},
  \bibinfo{pages}{1188–1191} (\bibinfo{year}{2010}).

\bibitem{Rokni2020}
\bibinfo{author}{Rokni, H.} \& \bibinfo{author}{Lu, W.}
\newblock \bibinfo{title}{Direct measurements of interfacial adhesion in {2D}
  materials and van der {W}aals heterostructures in ambient air}.
\newblock \emph{\bibinfo{journal}{Nat. Commun.}} \textbf{\bibinfo{volume}{11}},
  \bibinfo{pages}{5607} (\bibinfo{year}{2020}).

\bibitem{Ma2017}
\bibinfo{author}{Ma, X.} \emph{et~al.}
\newblock \bibinfo{title}{Capillary-force-assisted clean-stamp transfer of
  two-dimensional materials}.
\newblock \emph{\bibinfo{journal}{Nano Lett.}} \textbf{\bibinfo{volume}{17}},
  \bibinfo{pages}{6961–6967} (\bibinfo{year}{2017}).

\bibitem{Cai2022}
\bibinfo{author}{Cai, J.}, \bibinfo{author}{Chen, H.}, \bibinfo{author}{Ke, Y.}
  \& \bibinfo{author}{Deng, S.}
\newblock \bibinfo{title}{A capillary-force-assisted transfer for monolayer
  transition-metal-dichalcogenide crystals with high utilization}.
\newblock \emph{\bibinfo{journal}{ACS Nano}} \textbf{\bibinfo{volume}{16}},
  \bibinfo{pages}{15016–15025} (\bibinfo{year}{2022}).

\end{thebibliography}

\subsection*{Acknowledgments}
The authors would like to thank Dr Ruben Guis and Dr Gerard Verbiest for their support and discussions about the adhesion of 2D materials to mica. This work was supported by the Ministry of Education, Singapore under Research Centre of Excellence award to the Institute for Functional Intelligent Materials, I-FIM (project No. EDUNC-33-18-279-V12) awarded to K.S.N., under the Tier 3 program (MOE-MOET32024-0001) awarded to K.S.N., and by the National Research Foundation, Singapore under its AI Singapore Programme (AISG Award No: AISG3-RP-2022-028) awarded to K.S.N. This material was based upon work supported by the Air Force Office of Scientific Research and the Office of Naval Research Global under award number FA8655-21-1-7026 awarded to M.K., as well as under the Academic Research Fund Tier 2 (MOE-T2EP50122-0012) awarded to M.K. This work has been supported by the National Research Foundation, Singapore, under its NRF Competitive Research Programme (CRP) (NRF-CRP33-2025R-0001) awarded to M.L. This work has been supported by NRF Fellowship (NRFF) (NRF-NRFF16-2024-0011) awarded to A.I.B., and by A*STAR under its RIE2025 Manufacturing, Trade and Connectivity (MTC) Young Individual Research Grant (M23M7c0126) awarded to A.I.B.

\subsection*{Author contributions statement}
M.\v{S}., I.B. and A.I.B. initiated the project. K.S.N., M.K., A.I.B., I.B. and M.\v{S}. supervised the project. I.B., T.S., N.M., K.V., N.Z., A.C. and M.\v{S}. fabricated the samples. D.B., I.B., T.S., N.M., N.Z., K.V., Y.Y. and A.C. performed AFM measurements and sample characterisation. Y.Y. and M.L. performed high-resolution cAFM measurements. I.B. performed low-temperature magneto-transport measurements. I.B. and A.I.B. analysed transport measurement. D.L., T.S. and M.\v{S}. performed nanomechanical measurements. V.G. and M.\v{S}. performed XPS measurements. K.W. and T.T. provided hBN crystals. The paper was jointly written by all authors with a main contribution from M.\v{S}. All authors discussed the results and commented on the paper.

\subsection*{Competing interests statement}
The authors declare no competing interests.

\end{document}